\RequirePackage{fixltx2e}
\documentclass[a4paper]{isp}
\usepackage[american]{babel}
\usepackage{graphicx}
\usepackage{amsmath,amsthm,mathtools}
\usepackage{url}
\usepackage[capitalise,nameinlink]{cleveref}
\usepackage{xcolor}

\def\aj{AJ  }
\def\apj{ApJ\,  }
\def\apjl{ApJ Letters,  }
\def\apjs{ApJS  }

\def\mnras{MNRAS\,  }

\def\rmxaa{Revista Mexicana de Astronomia y Astrofisica}

\def\2F1{~_2F_1}
\def\1F1{~_1F_1}
\def\pFq{~_pF_q}
\def\sun{\hbox{$\odot$}}
\def\sun{\odot}
\def\h0units{\mathrm{km\,s^{-1}\,Mpc^{-1}}}
\newcommand{\om}{\Omega_{\rm M}}
\newcommand{\ola}{\Omega_{\Lambda}}
\begin{document}
\pdfgentounicode=1
\title
{
New probability distributions in astrophysics:
I. The   truncated generalized gamma.
}
\author{Lorenzo Zaninetti}
\institute{
Physics Department,
 via P.Giuria 1, I-10125 Turin,Italy \\
 \email{zaninetti@ph.unito.it}
}

\maketitle

\begin {abstract}
The gamma function is a good approximation to the luminosity
function of astrophysical objects,
and a truncated gamma distribution would permit a
more rigorous analysis.
This paper examines  the generalized gamma  distribution (GG)
and then introduces the scale and the new double
truncation.
The magnitude version of the truncated GG
distribution with scale is
adopted in order to fit  the luminosity function (LF) for galaxies or quasars. 
The new truncated GG LF  is applied 
to 
the five bands of SDSS galaxies,
to the 2dF QSO Redshift Survey
in the range of redshifts between 0.3 and 0.5,
and to the COSMOS QSOs 
in the range of redshifts between 3.7 and  4.7.
The average absolute magnitude versus redshifts for  
SDSS galaxies 
and 
QSOs of 2dF  
was modeled adopting  a redshift dependence for  
the lower and  upper absolute magnitude
of the new truncated  GG LF.
\end{abstract}

\section{Introduction}

The generalized gamma  distribution (GG) 
was introduced by \cite{Stacy1962} 
and carefully analyzed by  
\cite{Khodabina2010}.
The  GG has three-parameters and 
the techniques to find them
is a matter of research, see among others 
\cite{Stacy1965,Gomes2008,Vani2016}.
A significant role  in astrophysics is played  
by the  luminosity function (LF) for galaxies   and 
we present  some models,  among others,
the  Schechter LF, 
see \cite{schechter},
a two-component Schechter-like LF,
see \cite{Driver1996}, and 
the double Schechter LF  with five
parameters, see \cite{Blanton_2005}.
Another approach starts  from a given  
statistical distribution  for which  the probability 
density function (PDF) is known.
We know  that for a given PDF, $f(L)$, 
\begin{equation}
\int_0^{\infty} 
f(L) dL =1 
\quad ,
\end{equation} 
where $L$ is the luminosity.
A data oriented LF, $\Psi(L)$,  is obtained by 
adopting  $\Psi^*$ which is the normalization 
to the number of galaxies in a volume of 1 $Mpc^3$
\begin{equation}
\Psi(L) = \Psi^* f(L)
\quad ,
\end{equation}
which means 
\begin{equation}
\int_0^{\infty} 
\Psi(L) dL = \Psi^*
\quad .
\end{equation} 
The  above  line of research allows exploring  
the LF for galaxies in the framework 
of well studied PDFs.
Some examples are represented by 
the mass-luminosity relationship,
see \cite{Zaninetti2008},
some models connected  with the generalized gamma (GG) distribution,
see \cite{Zaninetti2010f,Zaninetti2014b},
the truncated beta  LF,
see  \cite{Zaninetti2014d},
the lognormal LF,
see \cite{Zaninetti2018d},
the truncated lognormal LF,
see \cite{Zaninetti2016c},
and the Lindley LF,
see \cite{Zaninetti2019a}.

This paper  brings up 
the GG and introduces the scale   
in Section \ref{secgammagen}. 
The  new double   truncation
for the GG  and the GG with scale is introduced 
in Section \ref{secgammagentronc}.
The derivation of the truncated GG  LF 
is done in Section \ref{secluminosity}.
Section \ref{secastro} contains the 
application  of the GG LF to galaxies and quasars
as well the fit of the averaged absolute magnitude 
for QSOs as function of the redshift.

\section{The generalized gamma distribution}

\label{secgammagen}
Let $X$ be a random variable
defined in
$[0, \infty]$;
the  GG (PDF), $f(x)$,
is
\begin{equation}
f (x;a,b,c) =
\frac
{
c{b}^{{\frac {a}{c}}}{x}^{a-1}{{\rm e}^{-b{x}^{c}}}
}
{
\Gamma \left( {\frac {a}{c}} \right) 
}
\quad ,
\end{equation}
where
\begin{equation}
\mathop{\Gamma\/}\nolimits\!\left(z\right)
=\int_{0}^{\infty}e^{{-t}}t^{{z-1}}dt
\quad ,
\end{equation}
is the gamma function,
with   $a>0$, $b>0$ and $c>0$.
The above PDF can be obtained by setting  
the location parameter equal to zero
in the four parameters GG
as  given by \cite{evans2011},pag.~113.
The   GG PDF scale as $\exp(-x^c)$  and the 
gamma PDF  as  $\exp(-x)$: the introduction of the parameter $c$
increases the flexibility of the GG.

The GG family  includes several  subfamilies, including 
the exponential PDF when $a=1$ and $c=1$, 
the gamma PDF when $c=1$ 
and the Weibull PDF when $b=1$ .

The  distribution function (DF), $F(x)$,
 is
\begin{equation}
F (x;a,b,c) =
1-
\frac
{
{\Gamma \left( {\frac {a}{c}},b{x}^{c} \right)}
}
{
{
\left( \Gamma
 \left( {\frac {a}{c}} \right)  \right)}
}
\quad,
\end{equation}
where $\Gamma(a, z)$ is the incomplete Gamma function,
defined by
\begin{equation}
\mathop{\Gamma\/}\nolimits\!\left(a,z\right)
=\int_{z}^{\infty}t^{a-1}e^{-t}dt
\quad ,
\end{equation}
see  \cite{NIST2010}.
The average  value or mean, $\mu$,  is
\begin{equation}
\mu (a,b,c)=
\frac
{
{b}^{-{c}^{-1}}\Gamma \left( {\frac {1+a}{c}} \right) 
}
{
\Gamma \left( {\frac {a}{c}} \right) 
}
\quad ,
\end{equation}
the variance, $\sigma^2$, is
\begin{equation}
\sigma^2(a,b,c)=
\frac
{
{b}^{-2\,{c}^{-1}} \left( - \left( \Gamma \left( {\frac {1+a}{c}}
 \right)  \right) ^{2}+\Gamma \left( {\frac {a}{c}} \right) \Gamma
 \left( {\frac {2+a}{c}} \right)  \right) 
}
{
\left( \Gamma \left( {\frac {a}{c}} \right)  \right) ^{2}
}
\quad .
\end{equation}
The mode, $M$,  is  
\begin{equation}
M(a,b,c)= \sqrt [c]{{\frac {a-1}{bc}}}
\quad .
\end{equation}
The rth moment about the origin is,
$\mu^{\prime}_r(a,b,c)$,
is
\begin{equation}
\mu^{\prime}_r(a,b,c) =
\frac
{
{b}^{-{\frac {r}{c}}}\Gamma \left( {\frac {a+r}{c}} \right) 
}
{
\Gamma \left( {\frac {a}{c}} \right)
}
\quad .
\end{equation}
The information entropy, $H$, 
is 
\begin{equation}
H(a,b,c)= 
\ln  \left( {\frac {1}{c\sqrt [c]{b}}\Gamma \left( {\frac {a}{c}}
\right) } \right) +\Psi \left( {\frac {a}{c}} \right)  \left( \frac{1}{c}-{\frac {a}{c}} \right) +{\frac {a}{c}}
\quad ,
\end{equation}
where $\Psi (z)$ is the digamma or Psi function  
defined as
\begin{equation}
\Psi\left(z\right)=\Gamma'\left(z\right)/\Gamma\left(z\right)
\quad ,
\end{equation} 
where $\mathcal{R}z > 0$, 
see  \cite{NIST2010}.

\subsection{The scale}

In some applications it may be useful to have a scale, $b$,
and therefore the  GG PDF, $f_s(x)$,
is
\begin{equation}
f_s (x;a,b,c) =
\frac
{
a \left( {\frac {x}{b}} \right) ^{ac-1}{{\rm e}^{- \left( {\frac {x}{b
}} \right) ^{a}}}
}
{
b\Gamma \left( c \right) 
}
\quad ,
\label{gammagenscale}
\end{equation}
which has DF, $F_s(x)$,
as 
\begin{equation}
F_s (x;a,b,c) =
\frac
{
{x}^{ac}{b}^{-ac}{{\rm e}^{-{{\rm e}^{a \left( \ln  \left( x \right) -
\ln  \left( b \right)  \right) }}}}+\Gamma \left( c \right) c-\Gamma
 \left( c+1,{{\rm e}^{a \left( \ln  \left( x \right) -\ln  \left( b
 \right)  \right) }} \right) 
}
{
\Gamma \left( c \right) c
}
\quad.
\end{equation}

The average  value, $\mu_s$,   is
\begin{equation}
\mu_s (a,b,c)=
\frac
{
b\Gamma \left( {\frac {ac+1}{a}} \right) 
}
{
\Gamma \left( c \right)
}
\quad ,
\end{equation}
the variance, $\sigma_s^2$, is
\begin{equation}
\sigma_s^2(a,b,c)=
\frac
{
{b}^{2} \left( \Gamma \left( c \right) \Gamma \left( {\frac {ac+2}{a}}
 \right) - \left( \Gamma \left( {\frac {ac+1}{a}} \right)  \right) ^{2
} \right) 
}
{
\left( \Gamma \left( c \right)  \right) ^{2}
}
\quad ,
\end{equation}
and 
the mode, $M_s$,  is  
\begin{equation}
M_s=\sqrt [a]{{\frac {ac-1}{a}}}b
\quad .
\end{equation}

\section{Double truncation}

\label{secgammagentronc}
Let $X$ be a random variable
defined in
$[x_l,x_u]$;
the new {\em double truncated } GG 
PDF, $f_t(x;a,b,c,x_{l},x_{u})$,
can be found 
by   evaluating  of the following integral
\begin{equation}
I_t(x;a,b,c) =
\int {x}^{a-1}{{\rm e}^{-b{x}^{c}}} dx 
\quad ,
\end{equation}
which is
\begin{eqnarray}
I_t(x;a,b,c) =
{\frac {{{\rm e}^{-1/2\,b{x}^{c}}}}{a   ( a+c   )    ( a+2\,
c   ) } \Bigg  ( \Big   ( {x}^{-3/2\,c+a/2}   ( a+c   ) {b}^{-
1/2\,{\frac {a+3\,c}{c}}}+c{x}^{-c/2+a/2}{b}^{-1/2\,{\frac {a+c}{c}}}
\Big   )
}
\times \nonumber  \\
{
 c{{\sl M}_{1/2\,{\frac {-c+a}{c}},\,1/2\,{\frac {a+2\,c}{c}}}
  (b{x}^{c}  )}+{b}^{-1/2\,{\frac {a+3\,c}{c}}}{{\sl M}_{1/2\,{
\frac {a+c}{c}},\,1/2\,{\frac {a+2\,c}{c}}}  (b{x}^{c}  )}{x}^{
-3/2\,c+a/2}   ( a+c   ) ^{2}  \Bigg ) }
\quad ,
\end{eqnarray}
where ${{\sl M}_{\mu,\,\nu}\left(z\right)}$ is 
the Whittaker $M$ function,  see Appendix \ref{appendix_whittaker}.
We now define the constant of integration, $K(a,b,c)$,
as
\begin{equation}
K(a,b,c,x_l,x_u) = \frac{1}
{
I_t(x_u;a,b,c) - I_t(x_l;a,b,c) 
}
\quad , 
\end{equation}
and  as a consequence the truncated GG  PDF is,
$f_t (x;a,b,c,x_l,x_u)$, 
\begin{equation}
f_t (x;a,b,c) = K(a,b,c,x_l,x_u)\times {x}^{a-1}{{\rm e}^{-b{x}^{c}}}
\quad .
\label{pdfggtruncated}
\end{equation}
The average  value, $\mu (a,b,c,x_l,x_u)$,   is 
\begin{eqnarray}
\mu (a,b,c,x_l,x_u)= K \times 
{\frac {{{\rm e}^{-1/2\,b{x}^{c}}}}{C} \big (    ( A+B   ) c{
{\sl M}_{1/2\,{\frac {-c+a+1}{c}},\,1/2\,{\frac {2\,c+a+1}{c}}}  (b
{x}^{c}  )}
}
\nonumber \\
{
+{{\sl M}_{1/2\,{\frac {c+a+1}{c}},\,1/2\,{\frac {2\,c+
a+1}{c}}}  (b{x}^{c}  )}{b}^{-1/2\,{\frac {3\,c+a+1}{c}}}{x}^{-
3/2\,c+a/2+1/2}D  \big ) }
\quad ,
\label{aveggtruncated}
\end{eqnarray}
where 
\begin{equation}
A ={x}^{-3/2\,c+a/2+1/2} \left( c+a+1 \right) {b}^{-1/2\,{\frac {3\,c+a+1
}{c}}}
\quad ,
\end{equation}
\begin{equation}
B=
c{x}^{a/2+1/2-c/2}{b}^{-1/2\,{\frac {c+a+1}{c}}}
\quad ,
\end{equation}
\begin{equation}
C=
\left( 2\,c+a+1 \right)  \left( c+a+1 \right)  \left( a+1 \right)
\quad ,
\end{equation}
\begin{equation}
D=\left( c+a+1 \right) ^{2}
\quad .
\end{equation}
The DF,  $DF(x;a,b,c,x_l,x_u)$, is
\begin{eqnarray}
DF(x;a,b,c,x_l,x_u)= K \times 
{\frac {{{\rm e}^{-1/2\,b{x}^{c}}}}{F} \big  ( {{\sl M}_{1/2\,{\frac {a
+c}{c}},\,1/2\,{\frac {a+2\,c}{c}}}  (b{x}^{c}  )}{b}^{-1/2\,{
\frac {a+3\,c}{c}}}{x}^{-3/2\,c+a/2}G
}
\nonumber \\
{
+E{{\sl M}_{1/2\,{\frac {-c+a}{c}
},\,1/2\,{\frac {a+2\,c}{c}}}  (b{x}^{c}  )}\,c  \big ) }
\quad ,
\label{dfggtruncated}
\end{eqnarray}
where 
\begin{equation}
E={x}^{-3/2\,c+a/2} \left( a+c \right) {b}^{-1/2\,{\frac {a+3\,c}{c}}}+c
{x}^{-c/2+a/2}{b}^{-1/2\,{\frac {a+c}{c}}}
\quad,
\end{equation}
\begin{equation}
F=a \left( a+c \right)  \left( a+2\,c \right) 
\quad,
\end{equation}
\begin{equation}
G= \left( a+c \right) ^{2}
\quad .
\end{equation}

\subsection{The scale}

The truncated GG  PDF with scale 
requires the evaluation of the following integral, $I_{ts}$,
\begin{equation}
I_{ts}(x;a,b,c) =
\int 
\left( {\frac {x}{b}} \right) ^{ca-1}{{\rm e}^{- \left( {\frac {x}{b}
} \right) ^{a}}}
dx 
\quad ,
\end{equation}
which is
\begin{eqnarray}
I_{ts}(x;a,b,c) =
\frac{1} {c   ( c+1
   ) a}
{b \Big  ( {x}^{1/2\,ac}{b}^{-1/2\,ac}{{\sl M}_{c/2,\,c/2+1/2}
  ({{\rm e}^{a   ( \ln    ( x   ) -\ln    ( b   ) 
   ) }}  )}
}
\nonumber \\
{
+{x}^{ac}{b}^{-ac}{{\rm e}^{-1/2\,{{\rm e}^{a
   ( \ln    ( x   ) -\ln    ( b   )    ) }}}}
   ( c+1   )  \Big  ) {{\rm e}^{-1/2\,{{\rm e}^{a   ( \ln 
   ( x   ) -\ln    ( b   )    ) }}}}}
\quad  .
\end{eqnarray}
The constant of integration, $K_s(a,b,c,x_l,x_u)$,
is
\begin{equation}
K_s(a,b,c,x_l,x_u) = \frac{1}
{
I_{ts}(x_u;a,b,c) - I_{ts}(x_l;a,b,c) 
}
\quad ,
\label{costanteks} 
\end{equation}
and  as a consequence the truncated GG  PDF with scale is,
$f_{ts} (x;a,b,c,x_l,x_u)$, 
\begin{equation}
f_{ts} (x;a,b,c,x_l,x_u) = K_s(a,b,c,x_l,x_u)\times 
\left( {\frac {x}{b}} \right) ^{ca-1}{{\rm e}^{- \left( {\frac {x}{b}
} \right) ^{a}}}
\quad .
\label{gammagenscaletronc}
\end{equation}
The average  value, $\mu_s (a,b,c,x_l,x_u)$   is 
\begin{eqnarray} 
\mu_s (a,b,c,x_l,x_u)= K_s \times 
{\frac {{{\rm e}^{-1/2\,{{\rm e}^{a   ( \ln    ( x   ) -\ln 
   ( b   )    ) }}}}}{C_s} \times
}
\nonumber \\
{
\bigg  ( a   ( A_s   ( 1+a
   ( c+1   )    ) {b}^{3/2+1/2\,   ( -c+3   ) a}+B_s
   ) {{\sl M}_{1/2\,{\frac {1+a   ( c-1   ) }{a}},\,1/2\,{
\frac {1+   ( c+2   ) a}{a}}}  ({{\rm e}^{a   ( \ln 
   ( x   ) -\ln    ( b   )    ) }}  )}
}
\nonumber \\
{
+A_s{b}^{3/2
+1/2\,   ( -c+3   ) a}{{\sl M}_{1/2\,{\frac {ac+a+1}{a}},\,1/2
\,{\frac {1+   ( c+2   ) a}{a}}}  ({{\rm e}^{a   ( \ln 
   ( x   ) -\ln    ( b   )    ) }}  )}D_s \bigg  ) 
}
\quad , 
\end{eqnarray}
where
\begin{equation}
A_s =
{x}^{1/2+1/2\,a \left( c-3 \right) }
\quad ,
\end{equation}
\begin{equation}
B_s =
a{x}^{1/2+1/2\,a \left( c-1 \right) }{b}^{3/2+1/2\, \left( -c+1
 \right) a}
\quad ,
\end{equation}
\begin{equation}
C_s =
\left( 1+a \left( c+1 \right)  \right)  \left( ac+1 \right)  \left( 1
+ \left( c+2 \right) a \right)
\quad ,
\end{equation}
\begin{equation}
D_s =
\left( 1+a \left( c+1 \right)  \right) ^{2}
\quad .
\end{equation}

\section{The luminosity function}

\label{secluminosity}
In this section 
we present the   Schechter LF,
we derive  the  GG LF,
we introduce the double truncation in the LF
and we develop  the adopted statistics.

\subsection{The  Schechter LF }

\label{secshechterlf}
The  Schechter LF,
introduced by
\cite{schechter},
is 
\begin{equation}
\Phi (L;\alpha,L^*,\Phi^*) dL  = (\frac {\Phi^*}{L^*}) (\frac {L}{L^*})^{\alpha}
\exp \bigl ( {-  \frac {L}{L^*}} \bigr ) dL \quad  ,
\label{equation_schechter}
\end {equation}
here $\alpha$ sets the slope for low values
of $L$, $L^*$ is the
characteristic luminosity and $\Phi^*$ is the normalization.
The luminosity density, $j$, is 
\begin{equation}
j= \int_0^{\infty} L\, \Phi (L;\alpha,L^*,\Phi^*) dL
\quad .
\label{j_schechter}
\end{equation}

The equivalent distribution in absolute magnitude is
\begin{equation}
\Phi (M)dM=0.921 \Phi^* 10^{0.4(\alpha +1 ) (M^*-M)}
\exp \bigl ({- 10^{0.4(M^*-M)}} \bigr)  dM \, ,
\label{lfstandard}
\end {equation}
where $M^*$ is the characteristic magnitude as derived from the
data.
We now  introduce  the parameter $h$
which is $H_0/100$, where $H_0$ is the Hubble constant.
The scaling with  $h$ is  $M^* - 5\log_{10}h$ and
$\Phi^* ~h^3~[Mpc^{-3}]$.

\subsection{Generalized gamma LF}

We replace in the GG  with scale, 
see equation~(\ref{gammagenscale})
$x$ with $L$   (the luminosity),
$b$ with $L^*$ (the characteristic luminosity) 
and we insert $\Psi^*$ which is the normalization 
to the number of galaxies in a volume of 1 $Mpc^3$
\begin{equation}
\Psi(L;a,b,c,\Psi^*) ={\it \Psi^*}
{\frac {a}{{\it L^*}\,\Gamma \left( c \right) } \left( 
{\frac {L}{{\it L^*}}} \right) ^{ac-1}{{\rm e}^{- \left( {\frac {L}{
{\it L^*}}} \right) ^{a}}}}
\quad . 
\end{equation}
The magnitude version is
\begin{equation}
\Psi(L;a,b,c,\Psi^*)
=
{\it \Psi^*}
\frac
{
0.4\,a{10}^{ac \left( - 0.4\,M+ 0.4\,{\it M^*} \right) }{{\rm e}^{
-{10}^{ \left( - 0.4\,M+ 0.4\,{\it M^*} \right) a}}}\,
\ln  \left( 10 \right) 
}
{
\Gamma \left( c \right)
}
\quad , 
\end{equation}
where $M$ is the absolute magnitude  
and   $M^*$ is the characteristic magnitude.
This four parameters LF, which was introduced in \cite{Zaninetti2010f},
was   applied to the  Sloan Digital Sky Survey (SDSS) in five different
bands
and  to the near infrared band 
of the 
2MASS Redshift Survey (2MRS),  see \cite{Zaninetti2014b}.

\subsection{Double truncation for the LF}

We replace in the truncated GG  with scale, 
see equation~(\ref{gammagenscaletronc}), 
$x$    with $L$   (the luminosity),
$b$    with $L^*$ (the characteristic luminosity),
$x_l$  with $L_l$ (lower luminosity),
$x_u$  with $L_u$ (upper luminosity), 
and we insert the  normalization, $\Psi^*$,
\begin{equation}
\Psi(L;a,L^*,c,L_l,L_u,\Psi^*)
= {\it \Psi^*} K_s(a,L^*,c,L_l,L_u)\times 
\left( {\frac {L}{L^*}} \right) ^{ca-1}{{\rm e}^{- \left( {\frac {L}{L^*}
} \right) ^{a}}}
\quad ,
\label{l_lfgammagentronc}
\end{equation}
where $K_s$ is given by equation (\ref{costanteks}).  
The luminosity density for the truncated GG  LF, $j_T$, 
has now a finite range of existence and is  
\begin{equation}
j_T= \int_{L_l}^{L_u} L\, \Psi(L;a,L^*,c,L_l,L_u,\Psi^*) dL
\quad .
\label{j_truncated}
\end{equation}

The four luminosities
$L,L_l,L^*$ and $L_u$
are  connected with  the
absolute magnitudes $M$,
$M_l$, $M_u$ and $M^*$
through the following relationship
\begin{equation}
\frac {L}{L_{\sun}} =
10^{0.4(M_{\sun} - M)}
\, ,
\frac {L_l}{L_{\sun}} =
10^{0.4(M_{\sun} - M_u)}
\,
, \frac {L^*}{L_{\sun}} =
10^{0.4(M_{\sun} - M^*)}
\,
, \frac {L_u}{L_{\sun}} =
10^{0.4(M_{\sun} - M_l)}
\label{magnitudes}
\end{equation}
where the indexes $u$ and $l$ are inverted in
the transformation
from luminosity to absolute magnitude
and $L_{\sun}$ and  $M_{\sun}$ are  the luminosity and absolute magnitude
of the sun in the considered band.

The magnitude version of the truncated GG  LF
is
\begin{equation}
\Psi(M;a,M^*,c,M_l,M_u,\Psi^*)
={\it \Psi^*}\frac{LD}{LN}
\label{lfgammagentronc}
\quad ,
\end{equation}
\begin{equation}
LD=0.4\,c \left( c+ 1  \right) a{{\rm e}^{-{{\rm e}^{ 0.921\,a
 \left( {\it M^*}-{\it M} \right) }}+ac \left(  0.921\,{
\it M^*}- 0.921\,{\it M} \right) }}\, \left( 
\ln  \left( 2 \right) +\ln  \left( 5 \right)  \right) 
\quad ,
\end{equation}
\begin{eqnarray}
LN=
{{\rm e}^{-{{\rm e}^{- 0.921\,a \left( {\it M_l}-{\it M^*}
 \right) }}+ \left(  0.921\,{\it M^*}- 0.921\,{\it 
M_l} \right) ca}}c-{{\rm e}^{-{{\rm e}^{ 0.921\,a \left( {\it 
M^*}-{\it M_u} \right) }}+ac \left(  0.921\,{\it M^*}-
 0.921\,{\it M_u} \right) }}c
\nonumber \\
+{{\rm e}^{- 0.5\,{{\rm e}^{-
 0.921\,a \left( {\it M_l}-{\it M^*} \right) }}+ \left( -
 0.46\,{\it M_l}+ 0.46\,{\it M^*} \right) ca}}{
{\sl M}_{c/2,\,c/2+1/2}\left({{\rm e}^{- 0.921\,a \left( {\it 
M_l}-{\it M^*} \right) }}\right)}
\nonumber \\
-{{\rm e}^{- 0.5\,{{\rm e}^{
 0.921\,a \left( {\it M^*}-{\it M_u} \right) }}+ \left( 
 0.46\,{\it M^*}- 0.46\,{\it M_u} \right) ca}}{
{\sl M}_{c/2,\,c/2+1/2}\left({{\rm e}^{ 0.921\,a \left( {\it 
M^*}-{\it M_u} \right) }}\right)}
\nonumber \\
+{{\rm e}^{-{{\rm e}^{-
 0.921\,a \left( {\it M_l}-{\it M^*} \right) }}+ \left( 
 0.921\,{\it M^*}- 0.921\,{\it M_l} \right) ca}}-{
{\rm e}^{-{{\rm e}^{ 0.921\,a \left( {\it M^*}-{\it M_u}
 \right) }}+ac \left(  0.921\,{\it M^*}- 0.921\,{\it 
M_u} \right) }}
\quad .
\end{eqnarray}

The averaged absolute magnitude   is  
\begin{equation}
{ \langle 
\Psi(M;a,M^*,c,M_l,M_u,\Psi^*) 
\rangle }
=
\frac{
\int_{M_l}^{M_u} 
\Psi(M;a,M^*,c,M_l,M_u,\Psi^*)
M dM
}
{
\int_{M_l}^{M_u} 
\Psi(M;a,M^*,c,M_l,M_u,\Psi^*)
dM 
}
\quad  .
\label{xmtruncatedgg4}
\end{equation}

\subsection{The adopted statistics}

The unknown parameters of the LF 
can be found 
through the Levenberg--Marquardt method (subroutine
MRQMIN in \cite{press}) but
the first derivative of the
LF with respect to the unknown parameters
should be provided.
In the  case  of the truncated GG  LF, see equation~(\ref{lfgammagentronc}),
the first derivative with respect to the unknown parameters 
has a complicated expression, so  we used 
the numerical first derivative.
The merit function $\chi^2$
can be computed by 
\begin{equation}
\chi^2 =
\sum_{j=1}^n ( \frac {LF_{theo} - LF_{astr} } {\sigma_{LF_{astr}}})^2
\quad ,
\label{chisquare}
\end{equation}
where $n$ is number of datapoints and the two 
 indexes $theo$ and $astr$ stand for theoretical 
and astronomical, respectively. 
The residual sum of squares (RSS) is
\begin{equation}
RSS =
\sum_{j=1}^n ( y(i)_{theo} -y(i)_{astr})^2
\quad ,
\label{rss}
\end{equation}
where  
$y(i)_{theo}$ is the theoretical value
and
$y(i)_{astr}$ is the astronomical value.
Particular attention should be paid to the number of  
unknown parameters
in the LF:  
three for the  Schechter function 
(formula~(\ref{lfstandard})) and 
four  for  formula~ (\ref{lfgammagentronc}).
The reduced  merit function $\chi_{red}^2$
can be computed by
\begin{equation}
\chi_{red}^2 = \chi^2/NF
\quad,
\label{chisquarereduced}
\end{equation}
where $NF=n-k$, $n$ being  the number of datapoints 
and $k$  the number of parameters.
The Akaike information criterion ($AIC$), see \cite{Akaike1974} ,
is defined by  
\begin{equation}
AIC  = 2k - 2  ln(L)
\quad ,
\end {equation}
where $L$ is
the likelihood  function  and $k$  the number of  free parameters
in the model.
We assume  a Gaussian distribution for  the errors
and  the likelihood  function
can be derived  by the $\chi^2$ statistic
$L \propto \exp (- \frac{\chi^2}{2} ) $
where  $\chi^2$ has been computed by  equation~(\ref{chisquare}),
see~\cite{Liddle2004}, \cite{Godlowski2005}.
Now $AIC$ becomes
\begin{equation}
AIC  = 2k + \chi^2
\quad.
\label{AIC}
\end {equation}

The Bayesian information criterion ($BIC$), see \cite{Schwarz1978},
is
\begin{equation}
BIC  = k~ ln(n) - 2  ln(L)
\quad,
\end {equation}
where $L$ is
the likelihood  function, 
$k$  the number of  free parameters
in the model and  $n$ the number of observations.
The phrase  "better fit" used in the following means that 
the three statistical indicators : $\chi^2$,
$AIC$ and $BIC$ are smaller for the considered LF 
than for the Schechter function .

\section{Astrophysical Applications}
\label{secastro}
In this section we apply the truncated GG LF 
to the SDSS galaxies  and to QSOs.
The introduction of the redshift dependence for  lower and upper absolute 
magnitude allows to model 
the average absolute magnitude versus redshift for QSOs.

\subsection{SDSS galaxies}

In order to perform a test  we selected the data 
of the  Sloan Digital Sky Survey (SDSS)    
which has  five bands  
$u^*$ ($\lambda = 3550\AA{})$,
$g^*$ ($\lambda = 4770\AA{})$,
$r^*$ ($\lambda = 6230\AA{})$,
$i^*$ ($\lambda = 7620\AA{})$
and 
$z^*$ ($\lambda = 9130\AA{})$
with  $\lambda$ denoting the wavelength of the CCD camera,
see \cite{Gunn1998}.
The data of the astronomical LF are reported in \cite{Blanton_2003}
and   are available at
\url{https://cosmo.nyu.edu/blanton/lf.html}.
The  numerical values of the four parameters $a$,
$c$, $M^*$
and $\Psi^*$  
are given in Table~\ref{datagene4tronc}.
 \begin{table}
 \caption[]
{
 Parameters for fits to LF
 in SDSS Galaxies with the 
 four parameters truncated GG LF  as represented by 
 formula (\ref{lfgammagentronc}).
}
 \label{datagene4tronc} 
 \[
 \begin{array}{lccccc}
 \hline
parameter    &  u^*   &  g^* & r^*  & i^* & z^*  \\ \noalign{\smallskip}
 \hline
 \noalign{\smallskip}
M_l - 5\log_{10}h   
 & -20.65 
 & -22.09  
 & -22.94  
 & -23.42  
 & -23.73
 \\
M_u - 5\log_{10}h   
 & -15.78 
 & -16.32  
 & -16.30  
 & -17.21 
 & -17.48 
 \\
M^* - 5\log_{10}h   
 & -17.34 
 & -19.45  
 & -20.28 
 & -20.29 
 & -20.77 
 \\
\Psi^* [h^3~Mpc^{-3}] 
 & 0.042
 & 0.043  
 & 0.052   
 & 0.038 
 & 0.042  
 \\
c 
 & 0.473 
 & 0.078  
 & 0.015   
 & 0.247  
 & 0.10   
 \\ 
a 
 & 0.842  
 & 1.02  
 & 0.942  
 & 0.839  
 & 0.866 
 \\ 
\chi^2 
 & 283.17 
 & 747.58 
 & 2185
 & 1867
 & 2916
\\
\chi_{red}^2 
 & 0.591 
 & 1.256
 & 3.261
 & 2.648
 & 3.961
 \\
AIC~k=4 
 & 291.17
 & 755.58 
 & 2193
 & 1874
 & 2923
\\
BIC~k=4 
 & 307.89
 & 773.16 
 & 2211
 & 1893
 & 2941
\\
\chi^2 Schechter
 & 330.73 
 & 753.3 
 & 2260
 & 2282 
 & 3245
\\
\chi_{red}^2 -Schechter 
 & 0.689
 & 1.263
 & 3.368
 & 3.232
 & 4.403
\\
\hline 
 \end{array}
 \]
 \end {table}

\begin{figure}
 \centering
\includegraphics[width=6cm,angle=-90]{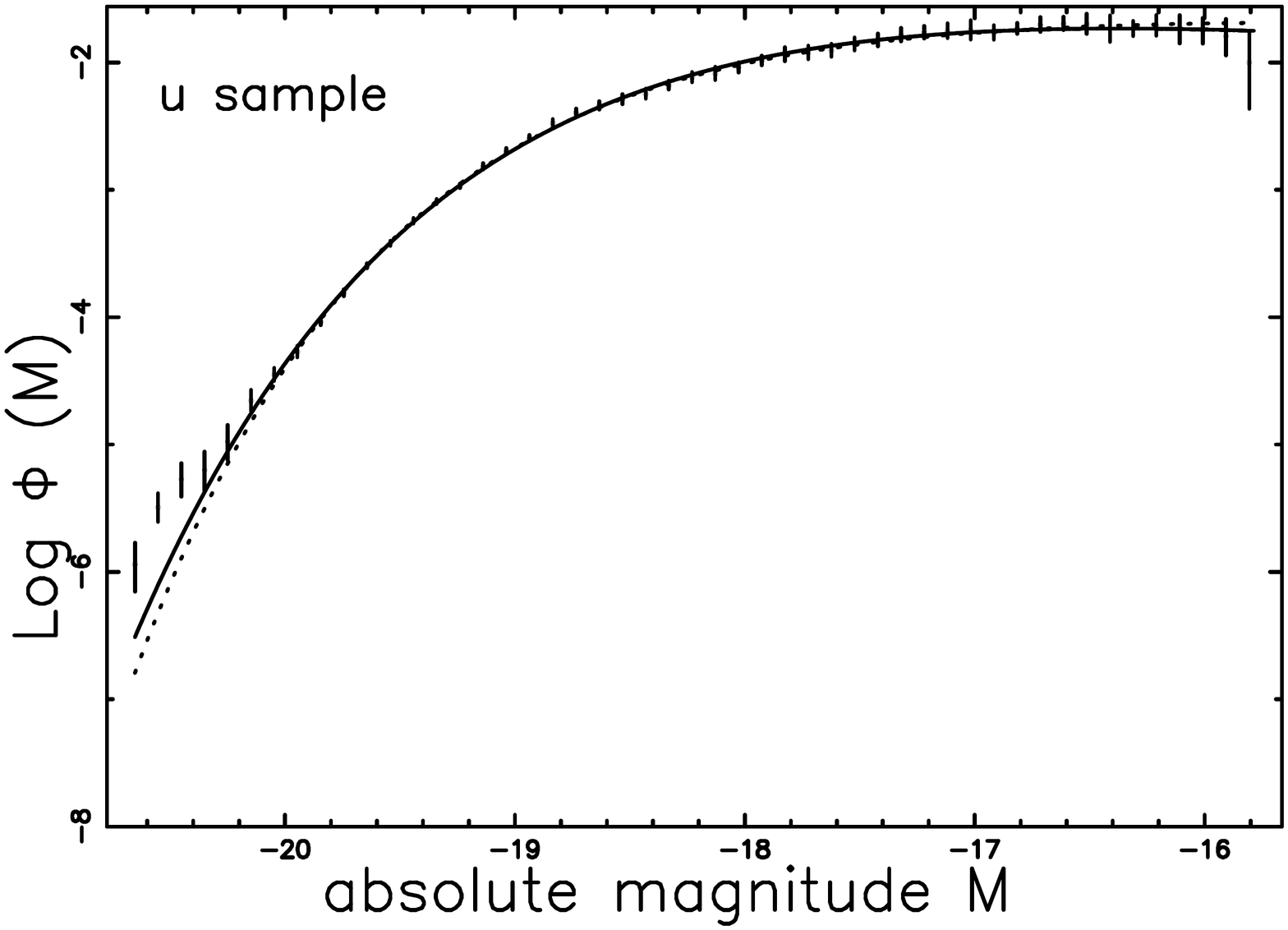}
\caption {The luminosity function data of 
SDSS($u^*$) are represented with error bars. 
The continuous line fit represents 
the four parameters truncated GG LF
(\ref{lfgammagentronc}) 
and the dotted 
line represents the Schechter function.
 }
 \label{due_u} 
 \end{figure}

\begin{figure}
 \centering
\includegraphics[width=6cm,angle=-90]{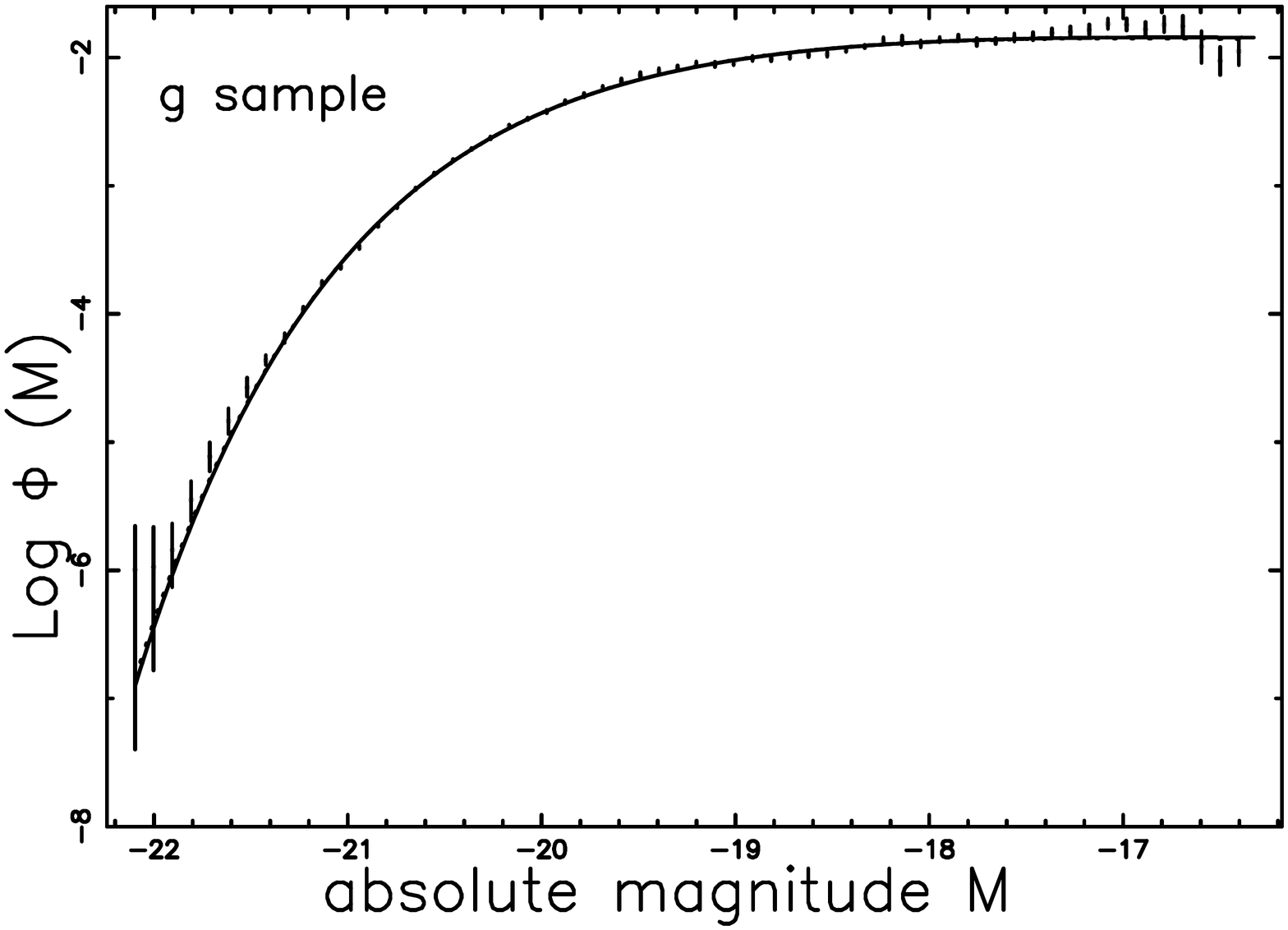}
\caption {The luminosity function data of 
SDSS($g^*$) are represented with error bars. 
The continuous line fit represents 
the four parameters truncated GG LF
(\ref{lfgammagentronc}) 
and the dotted 
line represents the Schechter function.
 }
 \label{due_g} 
 \end{figure}

\begin{figure}
 \centering
\includegraphics[width=6cm,angle=-90]{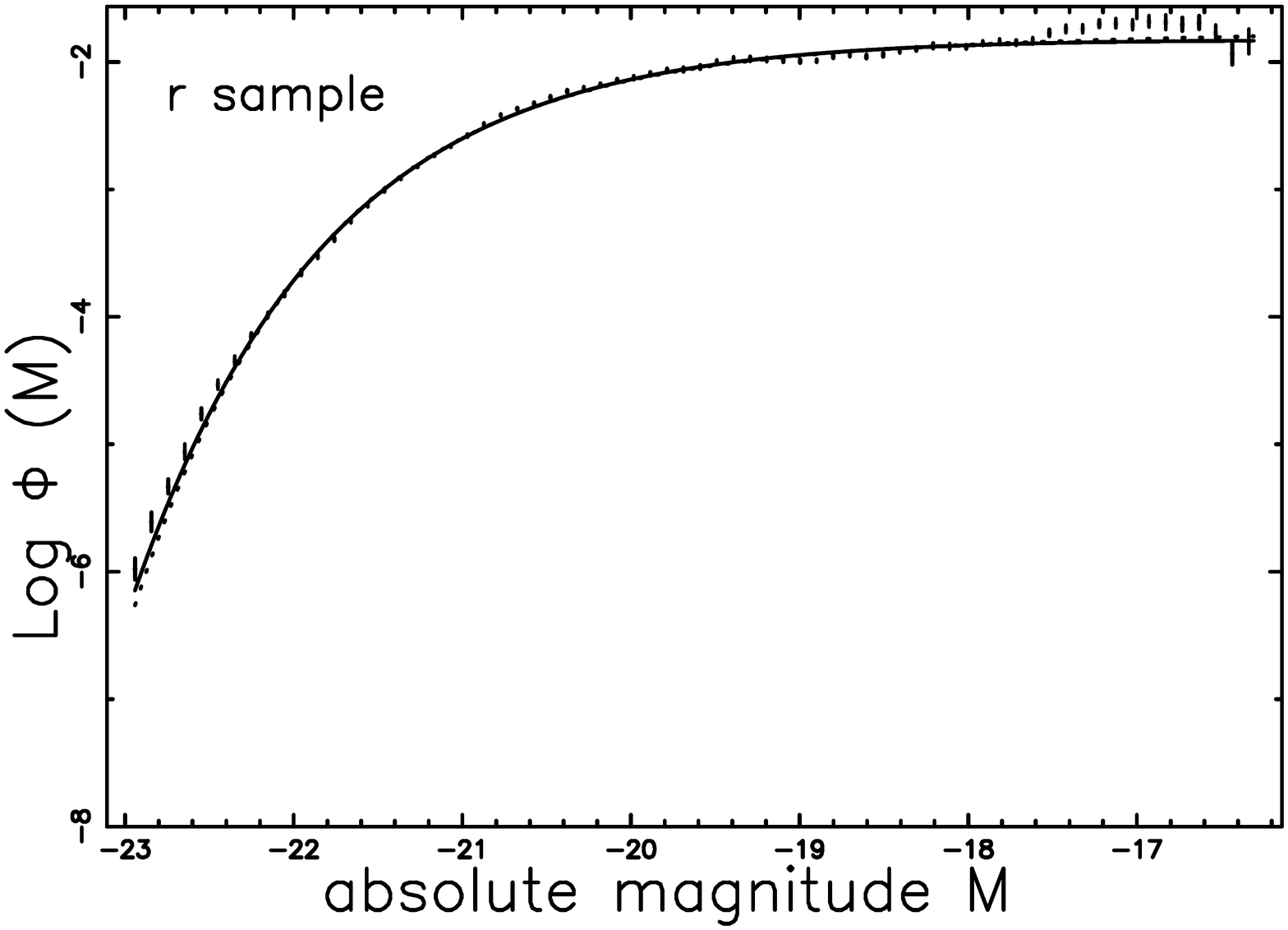}
\caption {The luminosity function data of 
SDSS($r^*$) are represented with error bars. 
The continuous line fit represents 
the four parameters truncated GG LF
(\ref{lfgammagentronc}) 
and the dotted 
line represents the Schechter function.
 }
 \label{due_r} 
 \end{figure}

\begin{figure}
 \centering
\includegraphics[width=6cm,angle=-90]{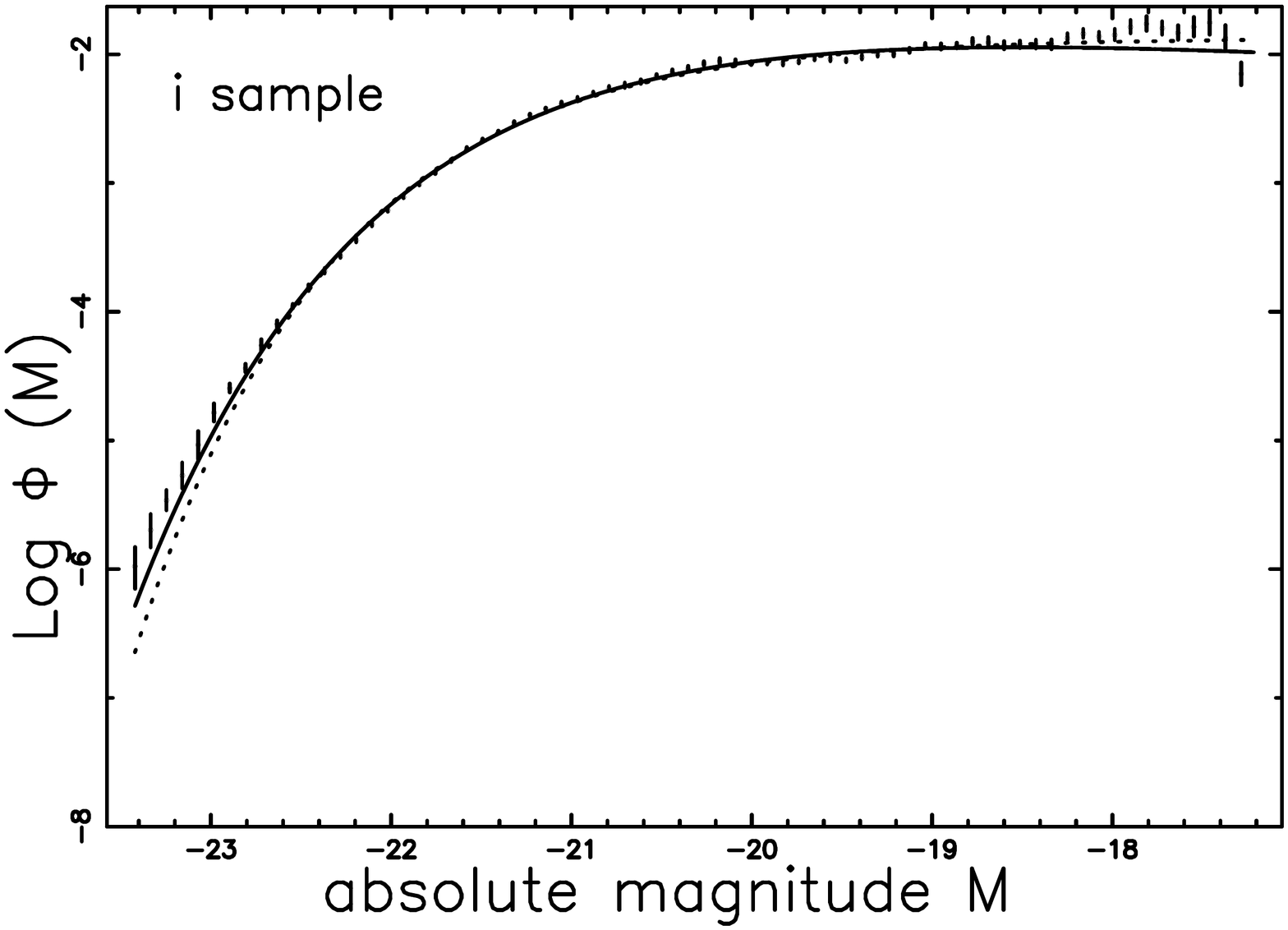}
\caption {
The luminosity function data of 
SDSS($i^*$) are represented with error bars. 
The continuous line fit represents 
the four parameters truncated GG LF
(\ref{lfgammagentronc}) 
and the dotted 
line represents the Schechter function.
 }
 \label{due_i} 
 \end{figure}

\begin{figure}
 \centering
\includegraphics[width=6cm,angle=-90]{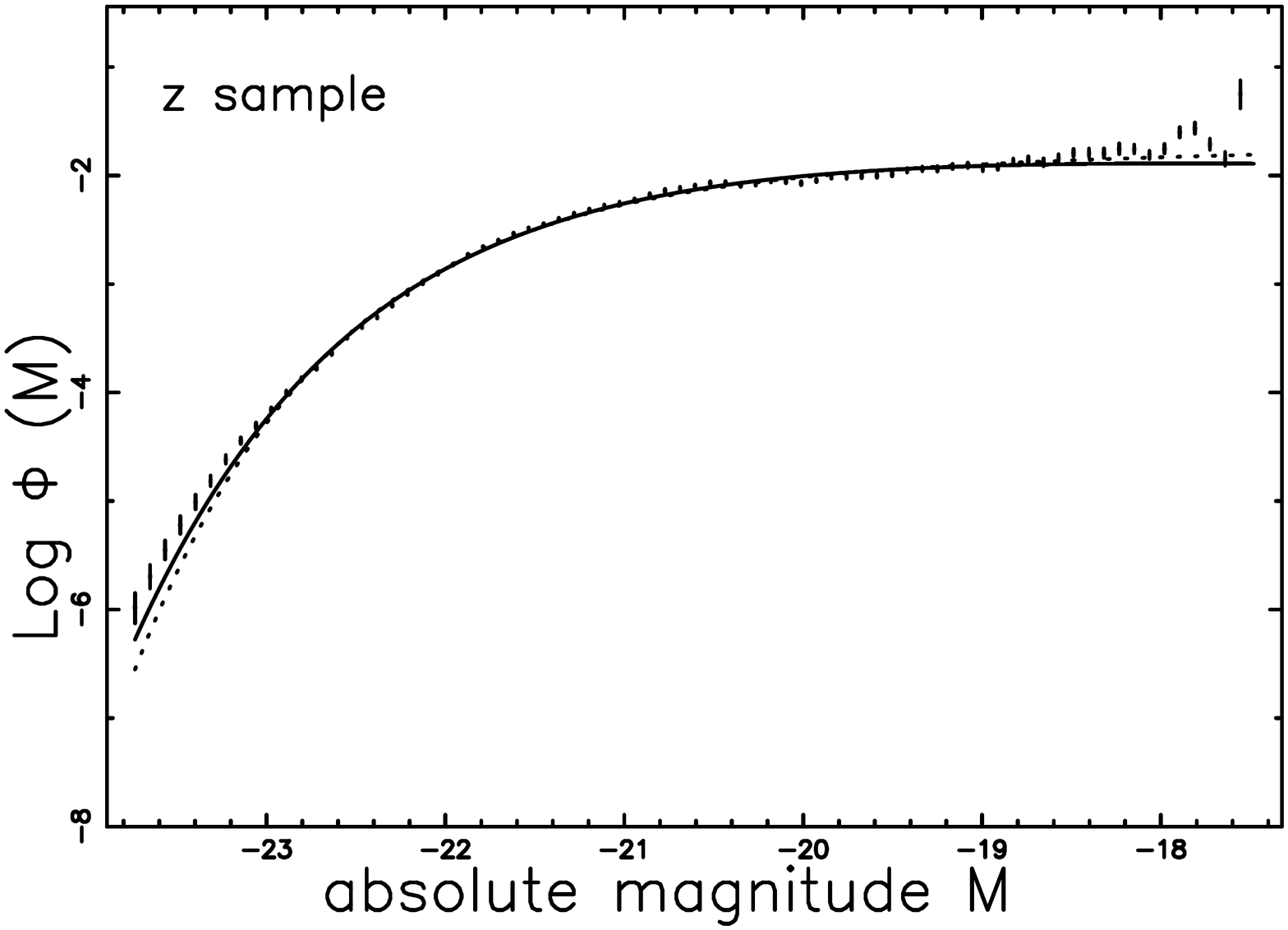}
\caption{
The luminosity function data of 
SDSS($z^*$) are represented with error bars. 
The continuous line fit represents 
the four parameters truncated GG LF
(\ref{lfgammagentronc}) 
and the dotted 
line represents the Schechter function.
 }
 \label{due_z} 
 \end{figure}

The Schechter function, the new four  parameters function 
as represented by formula~(\ref{lfgammagentronc})
and the data are
reported in 
Figure~\ref{due_u},
Figure~\ref{due_g},
Figure~\ref{due_r},
Figure~\ref{due_i} and 
Figure~\ref{due_z},
where bands 
$u^*$, $g^*$, $r^*$, $i^*$ and $z^*$ 
are considered.

Table \ref{dataj} presents the luminosity density 
evaluated with the Schechter LF, $j$,
and  with the truncated GG LF, $j_T$.
The range of existence in the truncated case 
is finite  rather than infinite and therefore 
the luminosity density is always smaller than in  the
standard case.

 \begin{table}
 \caption[]{
Luminosity density 
 in SDSS Galaxies 
evaluated with the
Schechter LF, formula (\ref{j_schechter}), 
and with the
 four parameters truncated GG LF, formula (\ref{j_truncated}),
when $\om=0.3$ and $\ola=0.7$. 
}
 \label{dataj} 
 \[
 \begin{array}{lccccc}
 \hline
parameter    &  u^*   &  g^* & r^*  & i^* & z^*  \\ \noalign{\smallskip}
 \hline
 \noalign{\smallskip}
j/(h\,10^8\,L_{\sun})  
 & 4.35
 & 2.81
 & 2.58
 & 3.19
 & 3.99
\\
j_T/(h\,10^8\,L_{\sun})  
 & 1.38
 & 1.18
 & 1.57
 & 1.88
 & 2.47
\\
\hline 
 \end{array}
 \]
 \end {table}

\subsection{Luminosity function for QSOs}

\begin{figure}
\begin{center}
\includegraphics[width=7cm,angle=-90]{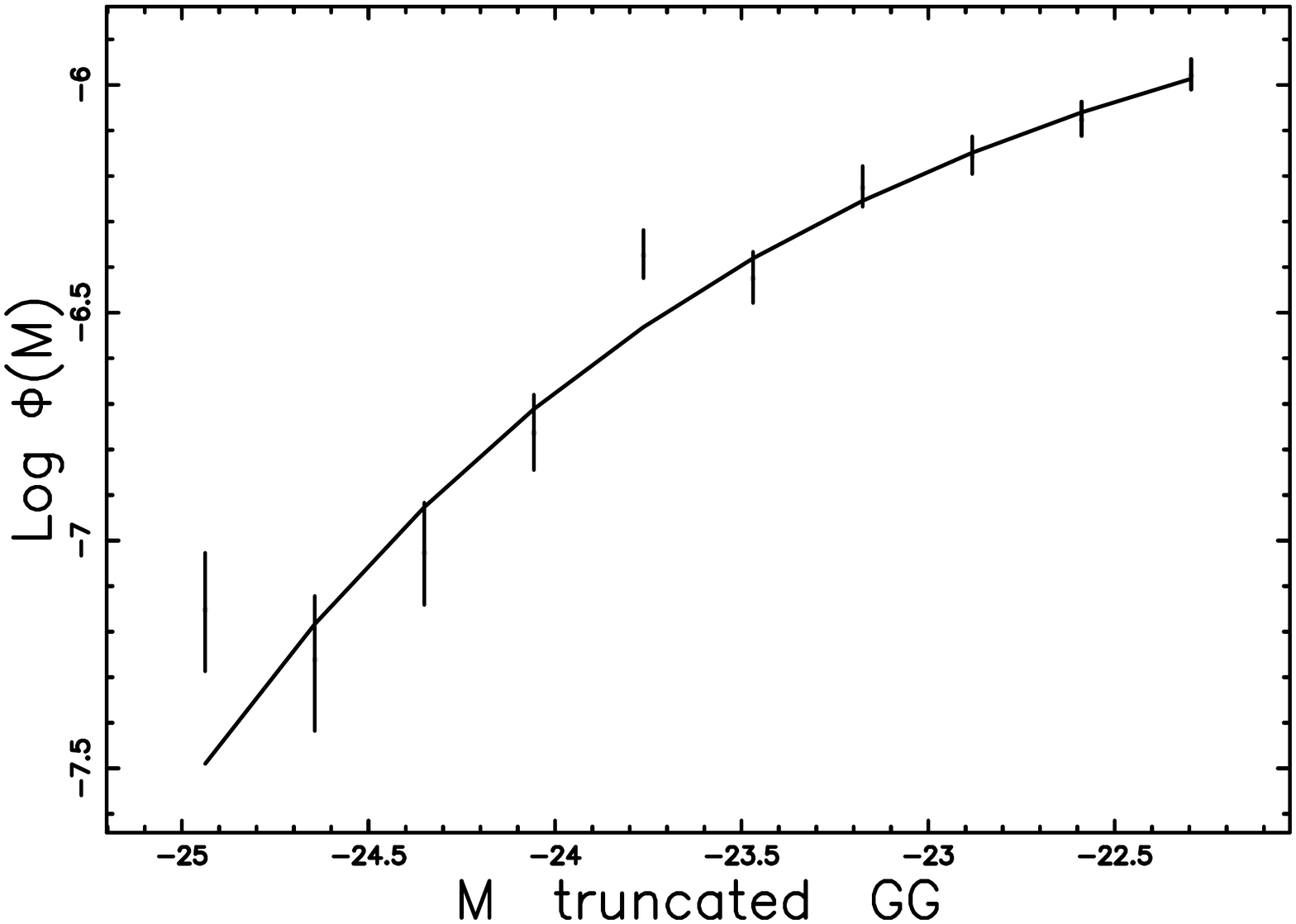}
\end{center}
\caption{
The  observed LF for QSOs, empty stars with error bar, 
when $z$ $[0.3,0.5]$
and  $M$ $[-24.93,-22.29 ]$.
The continuous line fit represents 
the four parameters truncated GG LF
(\ref{lfgammagentronc}) 
}
 \label{gg4qso}%
\end{figure}

For our  first example,
we selected the  catalog of the 2dF QSO Redshift Survey (2QZ),
which contains  22431 redshifts of QSOs  
with $18.25 <b_J< 20.85$, 
see \cite{Croom2004}.
We processed   them as  explained  in \cite{Zaninetti2017a}.
A typical  example of the observed LF for QSOs 
when  $ 0.3 < z< 0.5$ as well the  fit 
with the four parameters truncated GG LF
is presented
in Figure \ref{gg4qso}
with data as in Table~\ref{qsogene4tronc}.

 \begin{table}
 \caption[]{
The  four parameters truncated GG LF  as represented by 
 formula 
 (\ref{lfgammagentronc}) in the QSO case.
}
 \label{qsogene4tronc} 
 \[
 \begin{array}{lc}
 \hline
parameter    &  value  \\ \noalign{\smallskip}
 \hline
 \noalign{\smallskip}
M_l - 5\log_{10}h   
 & -24.93
 \\
M_u - 5\log_{10}h   
 &  -22.29 
 \\
M^* - 5\log_{10}h   
 & -22.48 
 \\
\Psi^* [h^3~Mpc^{-3}] 
 & 1.09\,10^{-6}
 \\
c 
 & 0.013
 \\ 
a 
 &  0.652  
 \\ 
\chi^2 
 & 10.17 
\\
\chi_{red}^2 
 & 1.69
 \\
AIC~k=4 
 & 18.17
\\
BIC~k=4 
 & 19.38
\\
\chi^2 Schechter
 & 10.49
\\
\chi_{red}^2 -Schechter 
 & 1.49
\\
\hline 
 \end{array}
 \]
 \end {table}

\begin{figure}
\begin{center}
\includegraphics[width=7cm,angle=-90]{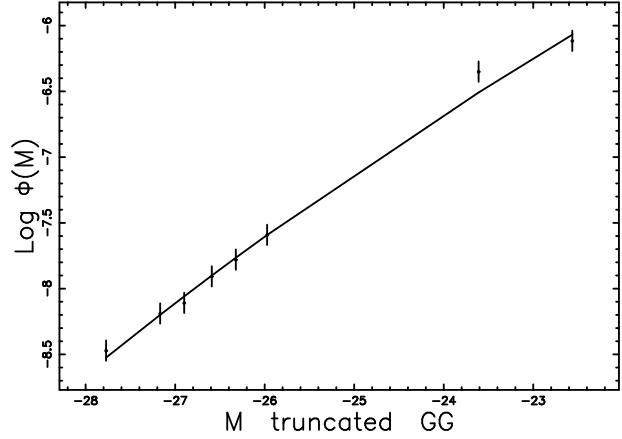}
\end{center}
\caption{
The  observed LF for QSOs in the COSMOS field, 
empty stars with error bar, 
when $z$ $[3.7,4.7]$
and  $M$ $[-28,-22 ]$.
The continuous line fit represents 
the four parameters truncated GG LF
(\ref{lfgammagentronc}) 
}
 \label{gg_4_cosmos_qso}%
\end{figure}

In the  second example 
we explored  the faint LF for quasars 
in the range of redshifts $3.7 < z < 4.7$ 
as  given in Figure 4 of \cite{Ikeda2011}.
The results are displayed 
in Figure \ref{gg_4_cosmos_qso}
with data as in Table~\ref{qsogene4tronc_cosmos}.

 \begin{table}
 \caption[]{
 The  
 four parameters truncated GG LF  as represented by 
 formula 
 (\ref{lfgammagentronc}) for QSOs in the COSMOS field.
}
 \label{qsogene4tronc_cosmos} 
 \[
 \begin{array}{lc}
 \hline
parameter    &  value  \\ \noalign{\smallskip}
 \hline
 \noalign{\smallskip}
M_l - 5\log_{10}h   
 & -25.86
 \\
M_u - 5\log_{10}h   
 &  -22.56 
 \\
M^* - 5\log_{10}h   
 &  -20.07
 \\
\Psi^* [h^3~Mpc^{-3}] 
 & 8.57\,10^{-7}
 \\
c 
 & -4.38
 \\ 
a 
 &  0.17  
 \\ 
\chi^2 
 &  3.46 
\\
\chi_{red}^2 
 & 0.86
 \\
AIC~k=4 
 & 11.46
\\
BIC~k=4 
 & 11.78
\\
\chi^2 Schechter
 & 5.82
\\
\chi_{red}^2 -Schechter 
 & 1.16
\\
\hline 
 \end{array}
 \]
 \end {table}

\subsection{Average absolute magnitude versus redshift}

\begin{figure}
\begin{center}
\includegraphics[width=10cm,angle=-90]{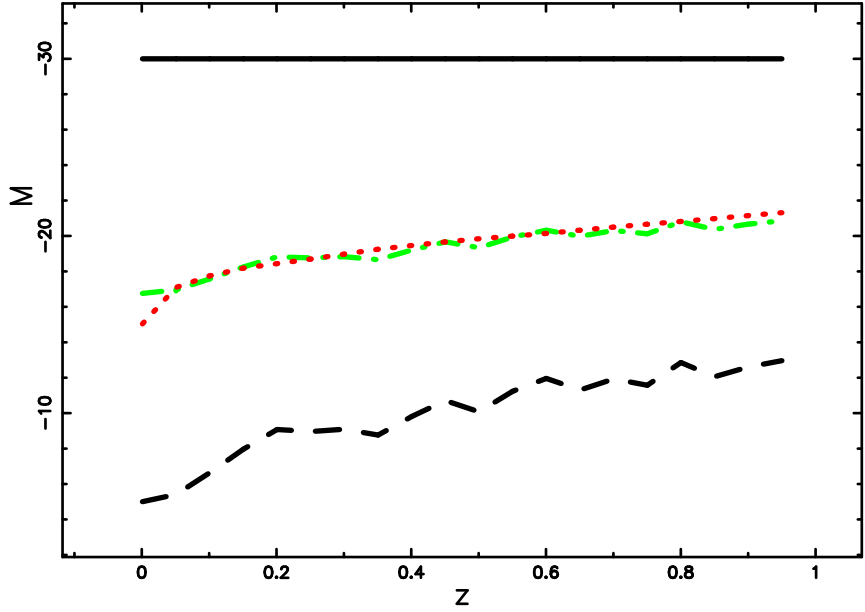}
\end{center}
\caption
{
Average observed  absolute magnitude 
versus redshift for SDSS galaxies  (red points),
average theoretical absolute magnitude 
for the four parameters of the truncated GG LF
as given by 
eqn~(\ref{xmtruncatedgg4})
(dot-dash-dot green line),
the lowest absolute magnitude  is $M=-30$ 
(full black line) and
the highest absolute magnitude  at  a given
redshift as  given by maximum value of the selected sample
(dashed black line);
RSS=5.14.
The others LF parameters for the truncated GG LF are:
$M^*=-29.8$, $a=0.1$ and $c=0.2$.
}
\label{lfsdss_avemag}
\end{figure}

The 
{\it first}  application is about 
galaxies:
we processed the  SDSS Photometric Catalogue DR 12,
see \cite{Alam2015}, which contains  10450256 galaxies 
(elliptical +
spiral)
with redshift and  rest frame $u^\prime$ absolute magnitude.
The lower  absolute magnitude  is fixed at  $M_l=-30$
and the upper absolute magnitude  is the 
maximum absolute magnitude   of the selected 
bin in redshift.
The above choice adopts the SDSS DR12 cosmological evaluation
of the absolute magnitude.
Figure \ref{lfsdss_avemag} displays 
averaged absolute magnitude, theoretical averaged absolute magnitude,
lower and upper limit in absolute magnitude  functions 
of the redshift.

The {\it second}  application is about the QSO 
and we used the framework   of the flat  cosmology
in order to find  the absolute magnitude  
relative to the catalog of the 2dF QSO Redshift Survey (2QZ),
exactly as \cite{Zaninetti2017a}.
The two cosmological  parameters are 
$\om=0.3$ and 
$H_0=70 \h0units$,
where $H_0$
is the Hubble constant expressed in     $\h0units$,
and  $\om$ is
\begin{equation}
\om = \frac{8\pi\,G\,\rho_0}{3\,H_0^2}
\quad ,
\end{equation}
where $G$ is the Newtonian gravitational constant and
$\rho_0$ is the mass density at the present time.
A useful  reference   for the upper magnitude 
as   function of  the redshift can be obtained from
the distance  modulus  as  given by equation (5) 
in \cite{Zaninetti2017a} once the limiting magnitude
of the sample  2QZ, $b_j=20.85$, is adopted  
\begin{eqnarray}
M_u =
\frac
{1}
{
\ln  \left( 2 \right) +\ln  \left( 5 \right) 
}
\Big (
- 4.15\,\ln    ( 2   ) - 4.15\,\ln    ( 5   ) +35\,\ln 
   (  1.0+z   ) -5\,\ln   \big  ( - 0.0032958754
\nonumber \\  
+
 10501.884\,{z}^{8}+ 64421.069\,{z}^{7}+
 167491.4963\,{z}^{6}+ 241951.366\,{z}^{5}+
 211426.1458\,{z}^{4}
\nonumber \\
+ 111997.68\,{z}^{3}+
 33297.7329\,{z}^{2}+ 4282.63\,z  \big  ) 
\Big )
\quad .
\end{eqnarray}
The above equation represents a useful theoretical reference.
Another, more empirical, way  explores  
numerically the maximum and the minimum  in absolute 
magnitude   functions of  the redshift  for the sample 2QZ.
In order  to fix  the numbers 
we fitted  the upper absolute magnitude  with the 
third degree polynomial 
\begin{equation}
M_u = a +b\,x +c\,x^2 +d\,x^3
\quad ,
\end{equation}
with   
$a=-17.75$, 
$b=-9.24$,
$c= 3.74$
and
$d=-0.58$.
The lower  absolute magnitude  is fitted with the
second degree polynomial
\begin{equation}
M_l = a +b\,x +c\,x^2 
\quad ,
\end{equation}
with   
$a=-24.86$, 
$b=-3.57$
and 
$c=0.45$.
Another useful relation
is  
\begin{equation}
M^*(z) = M_u(z)+0.1 
\label{mustarvariable}
\quad. 
\end{equation} 

\begin{figure}
\begin{center}
\includegraphics[width=10cm,angle=-90]{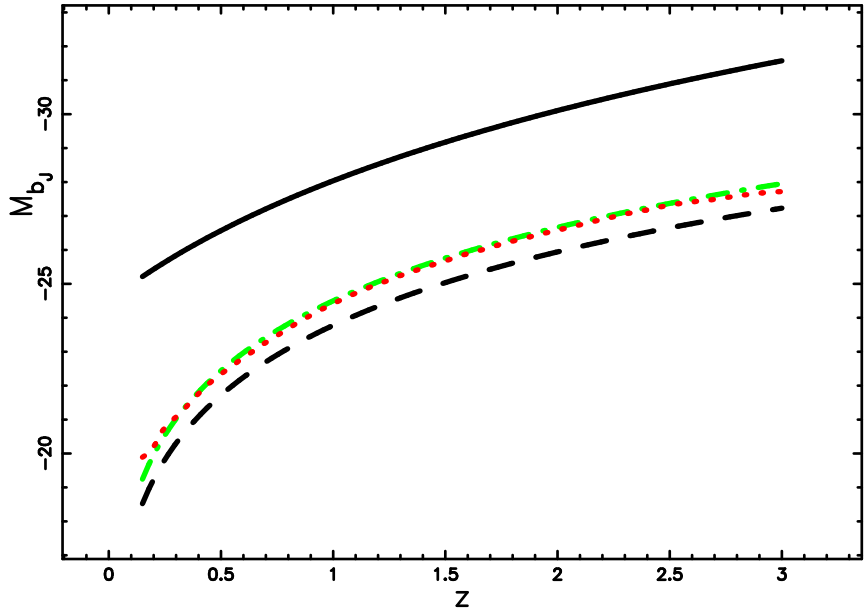}
\end{center}
\caption
{
Average observed  absolute magnitude 
versus redshift for QSOs  (red points),
average theoretical absolute magnitude 
for the four parameters of the truncated GG LF
as given by 
eqn~(\ref{xmtruncatedgg4})
(dot-dash-dot green line),
the lowest absolute magnitude  at  a given
redshift as  given by the sample
(full black line) and
the highest absolute magnitude  at  a given
redshift as  given by the sample
(dashed black line);
RSS=1.62.
}
\label{lfqso_avemag}
\end{figure}

Now different  combinations  of curves 
can be used.  Figure~\ref{lfqso_avemag} presents 
the combination
which minimizes the RSS.

\section{Conclusions}

{\bf Truncated GG:}  
We derived an expression for the left and right 
truncated GG PDF in terms  
of the Whittaker M function, 
see  equation (\ref{pdfggtruncated}),
its  DF, 
see  equation (\ref{dfggtruncated}),
and its  average value,
see  equation (\ref{aveggtruncated}).

\noindent
{\bf Truncated LF:}
The truncated LF for galaxies or QSO is derived
both in the luminosity form,
see equation (\ref{l_lfgammagentronc}),   
and in the magnitude form, 
see equation (\ref{lfgammagentronc}).
In all the  astrophysical examples here analyzed which are
the five bands of SDSS galaxies, 
see Table \ref{datagene4tronc},
the  QSOs
when  $ 0.3 < z< 0.5$,
see Table \ref{qsogene4tronc} ,
and
the faint LF for QSOs
in the range of redshift $3.7 < z < 4.7$, 
see Table \ref{qsogene4tronc_cosmos},
the $\chi^2$ of the truncated GG  is smaller
than  the Schechter LF.
Table \ref{dataj} presents  the 
luminosity density 
in SDSS Galaxies  with a finite range of existence rather
than infinite.

\noindent
{\bf Average Magnitude versus redshift:}  
The averaged absolute magnitude of the SDSS galaxies 
and  
QSOs  belonging to the catalog 2QZ as functions of the redshift 
are reasonably fitted by the  
averaged absolute magnitude  of the truncated GG LF,
see  equation~(\ref{xmtruncatedgg4}).
In order to perform the fit
we provided for $M^*$  a redshift dependence,
see equation (\ref{mustarvariable}),
and we inserted as lower and upper absolute magnitudes 
those given by the minimum and maximum values of the
selected bin in redshift.

\appendix
\setcounter{equation}{0}
\renewcommand{\theequation}{\thesection.\arabic{equation}}

\section{The Whittaker M function}

\label{appendix_whittaker}
The Whittaker M function,
${{\sl M}_{\mu,\,\nu}\left(z\right)}$,
solves the differential equation
\begin{equation}
{\frac {{\rm d}^{2}}{{\rm d}{z}^{2}}}W \left( z \right) + \left(
-\frac{1}{4}+
{\frac {\mu}{z}}+{\frac {\frac{1}{4}-{\nu}^{2}}{{z}^{2}}} \right) W \left( z
 \right) =0
\quad .
\end{equation} 
The   regularized hypergeometric
function,
 ${\2F1(a,b;\,c;\,z)}$, 
as defined by the Gauss series, is  
\begin{eqnarray}
\2F1(a,b;\,c;\,z)=
\sum_{s=0}^{\infty}\frac{{\left(a\right)_{s}}
{\left(b\right)_{s}}}{{\left(c\right)_{s}}s!}z^{s}
=1+\frac{ab}{c}z+\frac{a(a+1)b(b+1)}{c(c+1)2!}z^{2}+\cdots
\nonumber \\
=\frac{\Gamma\left(c\right)}{\Gamma\left(a\right)\Gamma
\left(b\right)}\sum_{s=0}^{\infty}\frac{\Gamma\left(a+s\right)\Gamma\left(b+s
\right)}{\Gamma\left(c+s\right)s!}z^{s}
\end{eqnarray} 
where
$z=x+iy$, 
and 
$(a)_s$
is the Pochhammer symbol
\begin{equation}
(a)_s=a(a+1)\dots(a+s-1)
\quad  .
\end{equation}
The generalized hypergeometric series 
is denoted by $\pFq$ and the case 
$p=1$ and $q=1$ 
gives 
\begin{eqnarray}
\1F1(a,b;\,z)=
\sum_{s=0}^{\infty}\frac{{\left(a\right)_{s}}
{ }}{{\left(b\right)_{s}}s!}z^{s}
\quad .
\end{eqnarray}

The relationship 
\begin{equation} 
{{\sl M}_{\mu,\,\nu}\left(z\right)}=
\frac
{
{z}^{\nu+\frac{1}{2}}{\mbox{$_1$F$_1$}
(-\mu+\nu+\frac{1}{2};\,1+2\,\nu;\,z)}
}
{
{{\rm e}^{\frac{z}{2}}}
}
\end{equation}
allows to express the Whittaker M function in terms
of the generalized hypergeometric function, $\1F1$,
see  \cite{Abramowitz1965,NIST2010}.
\providecommand{\newblock}{}

\end{document}